\documentclass[aps,twocolumn,groupedaddress, showpacs,prb]{revtex4}
\usepackage{graphicx}%

\begin{document}
\bibliographystyle{apsrev}


\title{Elements, Topology, and T-shirts}


\author{P. Fraundorf}
\affiliation{Physics \& Astronomy, U. Missouri-StL (63121), St. Louis, MO, USA}
\email[]{pfraundorf@umsl.edu}


\date{\today}

\begin{abstract}

Consider the network of elements connected by adjacency 
in atomic number and periodic table column.  
If all elements except H-He are connected by 
identical springs and constrained to a z-spiral in 2D, a map 
{\em sans} spring-crossing results for various values of 
rest length in units of the H-He distance.  If the 
z-spiral is removed, and the network 
allowed to relax in 3D, the resulting surface 
with three holes has the topology of a T-shirt or open-handle 
teapot.  Mapping this to a shirt's front-back pattern 
yields a 2D table of elemental 
inter-connects in tabular (nearly rectilinear) form.  
{\em Sample applications}: The flat spiral (like other 
spiral patterns) might serve as basis for an 
educational board game, while the T-shirt table might 
serve as a starting point for ``topologically informed'' 
classroom periodic-table design projects with wearable 
awards.
\end{abstract}
\maketitle

\tableofcontents
\section{Introduction}

The traditional periodic table shows elemental columns 
nicely, but only intermittently shows element connectivity 
via adjacent proton number (e.g. Ne-Na or Ba-La).  Hence
a variety of ``spiral'' and ``start-column'' variations 
have emerged \cite{mazurs74, katz01, cronyn03, rich05}, 
involving designer decisions on 
where to put things.  In this note, we explore 
ways to ``let the elements decide'' where to go, as if they 
are e.g. connected by a network of springs.  This 
is a standard ploy for visualizing networks.  The exploration 
also turned up an interesting topological 
connection between the periodic table and an item of 
everyday apparel.

\section{Springs on a spiral}

Begin by considering elements as movable beads-on-a-track, connected 
with springs when they are adjacent in atomic number and/or in 
traditional periodic table column assignment.  Our first reasonable pattern 
emerged with hydrogen, the noble gases, and elements heavier than xenon, fixed while letting the other elements slide along a spiral track under the influence of equal strength springs.  Further letting the outer elements slide along the spring also worked, but only if one weakens a subset of the springs e.g. for atoms beyond Argon.  Going back to equal strength springs, and allowing all elements except hydrogen and helium to go where they want on the wire, yields a slightly wilder layout but nonetheless without spring overlap for select values of ``spring rest length'' in units of the H-He distance.

\begin{figure}
\includegraphics[scale=.5]{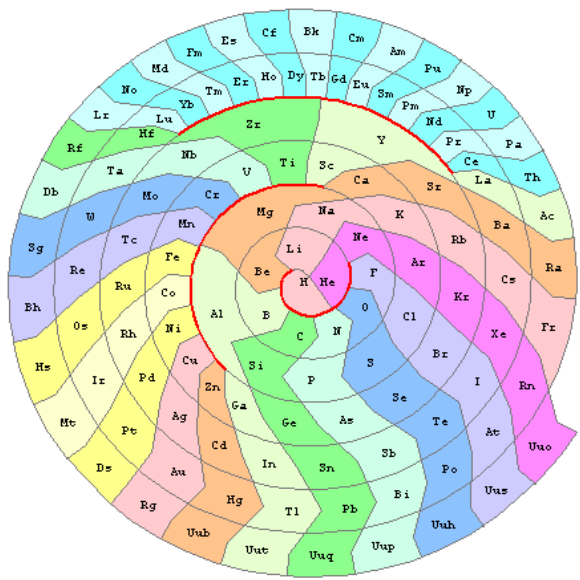}%
\caption{Here the spring rest length is set at 
2.5 times the fixed H-He distance.}
\label{ptspiral}
\end{figure}

The table form of this pattern, rotated to match the modern periodic table with noble gases on the right, is shown in Fig. \ref{ptspiral}. Given positions for hydrogen, helium (specifying scale and orientation) and lithium (specifying spring constant), all the other elements have decided where to go on their own. The patches for each element were then defined with an algorithm for locating intermediate vertices on spiral paths of the wire, and its 180 degree complement (which serves here to define the one continuous periodic table ``row''). As you might expect, this table has fifty elements in eight long (6 or 7 entry) "main-element" columns on the lower right, forty elements in ten intermediate (4-entry) ``transition-metal'' columns, and twenty-eight elements in fourteen short (2-entry) ``rare-earth'' columns at the top.

\section{Relaxing to three dimensions}

\begin{figure}
\includegraphics[scale=.65]{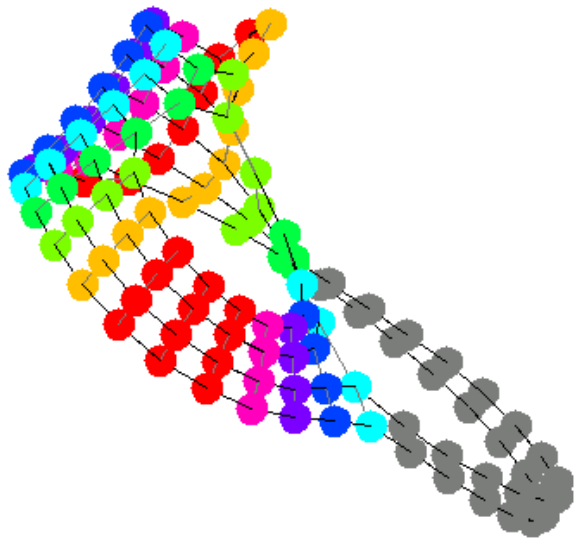}%
\caption{This network is relaxed using 
Mathematica's spring-electrical model.}
\label{pt3d}
\end{figure}

These connections between elements also rearrange themselves nicely under 
tension in three dimensions. In Fig. \ref{pt3d} we use Mathematica's 
spring-electrical model with adjacent element 
spring attraction, and a global ``one over r-squared'' repulsive force.  
In this screen capture from an interactive applet, note the black spiral pattern of increasing atomic number, as well as the three holes in the topographic sheet.  These holes are associated with three network bifurcations, and in effect give the 
sheet the same topology as an open-handled teapot (top, spout, handle) or a 
T-shirt (neck opening and two sleeves).

\section{T-shirt table of connections}

\begin{figure}
\includegraphics[scale=.42]{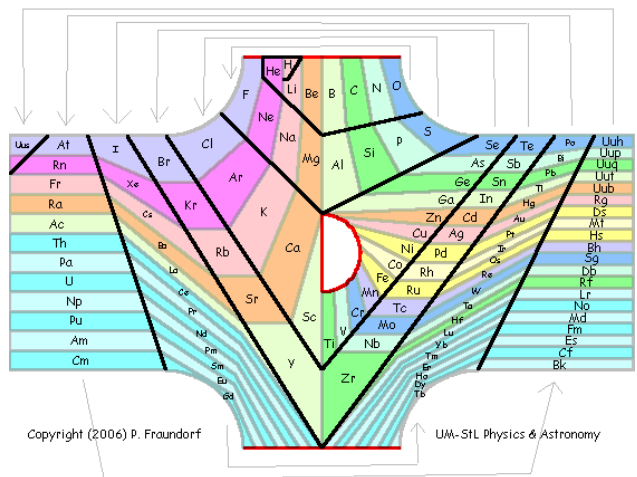}%
\caption{Here the mapping to T-shirt 
topology is designed to make the layout 
as rectilinear as possible.}
\label{ptshirt}
\end{figure}

The T-shirt topology suggests a way to display all connections in tabular (nearly rectilinear) form. Thus in Fig. \ref{ptshirt} a black line charts the path of the atomic-number spiral, while all columns of elements flow from one of three red-bordered (bifurcation) openings. The main element groups pour in from the top, transition metal groups from the central opening, and rare earth groups from the bottom.  One advantage to this format is that questions of period alignment \cite{katz01} become mute since the z-spiral is shown intact in this layout (as well as on the T-shirt).

Although the topology of the connections above is dictated by quantum mechanics, the shape and color of each region leaves quite a bit of room for artistic license to address problems like that of flow, color balance, information content, and ``crowding for the rare earths''. In fact, one might imagine ``periodic table shirt'' design contests that honor the topological constraints, but go for a variety of effects (both aesthetic and conceptual). They might even work in a junior high school math class, making for a nicely cross-disciplinary experience. How to manufacture the contest prize, i.e. a shirt without obtrusive blank spots, remains an open question.

\begin{acknowledgments}
Thanks to colleagues Chuck Granger, Hal Harris, and Keith Stine for useful suggestions. 
\end{acknowledgments}


\bibliography{ifzx.bib}

\end{document}